\documentclass[prd, aps, superscriptaddress, preprintnumbers, floatfix, nofootinbib]{revtex4}
\usepackage{amsfonts}
\usepackage{amsmath}
\usepackage{amssymb}
\usepackage{bm}
\usepackage{dcolumn}
\usepackage{graphicx}   
\usepackage[latin1]{inputenc}
\usepackage{latexsym}
\usepackage{rotating}
\usepackage{hyperref}
\usepackage{graphicx}
\usepackage{color}

\usepackage{ulem}

\newcommand\be{\begin{equation}}
\newcommand\ba{\begin{eqnarray}}
\newcommand\ee{\end{equation}}
\newcommand\ea{\end{eqnarray}}

\begin{document}

\title {Towards a Dark Sector Model from String Theory}

\author{Heliudson Bernardo}
\email{heliudson.deoliveirabernardo@mcgill.ca}
\affiliation{Department of Physics, McGill University, Montr\'{e}al, QC, H3A 2T8, Canada}

\author{Robert Brandenberger}
\email{rhb@physics.mcgill.ca}
\affiliation{Department of Physics, McGill University, Montr\'{e}al, QC, H3A 2T8, Canada and
Institute of Theoretical Physics, ETH Z\"urich, CH-8093 Z\"urich, Switzerland}

\author{J\"urg Fr\"ohlich}
\email{juerg@phys.ethz.ch}
\affiliation{Institute of Theoretical Physics, ETH Z\"urich, CH-8093 Z\"urich, Switzerland}

\date{\today}

\begin{abstract}

An embedding of a unified dark sector model into string theory with the following features is proposed: The model-independent axion descending from the Kalb-Ramond 2-form field is identified with the dark-matter field, and the real part of a K\"ahler modulus field -- the ``radius'' of one of the extra spatial dimensions -- accounts for dark energy. The expectation value of the dilaton field is stabilized by a gaugino condensation mechanism. A dark-energy potential corresponding to a realistic low energy scale results from some gentle tuning of the stabilized expectation value of the dilaton. The resulting potential reproduces the one in a previous dark-sector model proposed by two of us.

\end{abstract}

\pacs{98.80.Cq}
\maketitle

\section{Introduction} 
\label{sec:intro}

The origin of the dark sector, Dark Energy and Dark Matter, remains a mystery. The idea that degrees of freedom related 
to Quintessence \cite{Wetterich} may provide an explanation of the accelerated expansion of the universe has recently
inspired a proposal \cite{us} of a simple unified dark sector model involving a complex scalar field, $z= \sigma +i\theta$. 
The action functional of this field contains a self-interaction potential of the form
\be \label{toy}
V(\sigma, \theta) \, \simeq \, \bigl[ \Lambda + \frac{1}{2} \mu^4 {\rm{sin}}^2 (\theta / f) \bigr] e^{-2 \sigma / f} \,.
\ee
where $f$ is an energy scale that will be identified with the Planck mass, $m_{pl}$. 
The fields $\sigma$ and $\theta$ have canonical kinetic terms. In the proposal of \cite{us}, 
the field $\sigma$ is a source of Dark Energy, while oscillations of the axion field $\theta$ about a minimum of its potential yield 
Dark Matter \footnote{Note that the suggestion that an axion field could yield Dark matter was first made in \cite{DFSZ}.}. In \cite{us}, the dynamics of this model has been studied for initial conditions with $|\sigma/ f| \sim 1$ 
and $\theta$ frozen at a value $\theta \simeq f$, at an initial time corresponding to some critical temperature, $T_c$, 
higher than the temperature at the time of nucleosynthesis. If the constants $\Lambda$ and $\mu$ appearing 
in \eqref{toy} are chosen appropriately the second term (i.e., the Dark Matter term) in the potential dominates at early times, 
with $\theta$ exhibiting damped oscillations about a minimum of its potential. But once the contribution of the  
Dark Matter field $\theta$ has redshifted sufficiently, the first term (i.e., the Dark Energy term) starts to dominate, 
and the dynamics of the fields of the model mimics Quintessence.\footnote{See also \cite{us2} for related earlier attempts to construct a unified dark sector model, and \cite{Elisa} for a unified dark sector model in the context of superfluids.}

At the level of classical field theory various models with extra (classical, discrete or non-commutative) dimensions 
involve quite naturally a scalar field $\sigma$ with an \textit{exponential} self-interaction potential. 
This field could then provide a source of dynamical Dark Energy. In applications of such models 
to cosmology one usually uses them as \textit{classical field theories.} 
However, for consistency reasons, one should try to make sense of them as \textit{quantum field theories,} too.
Unfortunately though, after quantization a model field theory with an exponential self-interaction potential, as in 
Eq.~\eqref{toy}, (and with a potential periodic in the axion field) is not renormalizable perturbatively. One must therefore attempt to find an ultraviolet completion 
of the model. This leads one quite naturally to trying to embed it into string theory. 
Thus it is tempting to search for a superstring-theory origin of the model proposed in \cite{us}. 
In pursuing this idea one has to cope with a number of difficulties. To mention one, 
in most studies of superstring theory, the geometry of  the internal manifold of extra compact 
spatial dimensions in the ten-dimensional space-time is classical and static. 
It is natural to imagine that the scalar field $\sigma$ appearing in \eqref{toy} is related 
to some modulus of the internal manifold. Since $\sigma$ is dynamical 
and changes with cosmological time, one is apparently led to study the problem of making sense of string theories 
with \textit{dynamical} internal manifolds and a curved non-compact four-dimensional space-time. The study of such 
string theories has not reached a satisfactory stage, yet. Quite generally, our understanding of string cosmology
appears to be rather rudimentary.

In spite of such difficulties we propose a concrete embedding of the model described in \cite{us} into 
superstring theory. We choose to work in the framework of \textit{heterotic superstrings.} 
We consider a target space-time that is the
product of a four-dimensional non-compact Lorentzian space-time with a six-dimensional compact 
internal space given by some complex manifold\footnote{To leading order in the parameter $\alpha'$, 
the internal space is a K\"ahler manifold.} whose size may, however, depend on cosmological time.
In our embedding, the axion of the model-independent axio-dilaton superfield, $S$, 
is a source of Dark Matter, while the real part of  a K\"ahler volume-modulus superfield, 
$T$,  describing the size (extension) of one of the extra dimensions, is related to the Dark-Energy field $\sigma$. It is known that if no explicit volume-modulus 
stabilization mechanism is introduced the real part, $\text{Re} T$, of $T$ is a ``runaway'' direction. Frequently, 
stabilization mechanisms are constructed in order to pin the expectation value of this modulus field;
see, e.g., \cite{Baumann} for a review written for cosmologists. While such stabilization mechanisms 
can generate a local minimum in the potential of $\sigma$, the fact remains that the relevant minimum 
in field configuration space corresponds to decompactified extra dimensions. In fact, it requires fine-tuning 
of initial conditions for the evolution of $\sigma$ to start near a local minimum of its potential, 
and its dynamical evolution is expected to typically lead to ``overshooting'' of the local minimum, a phenomenon 
discussed a long time ago in \cite{Brustein}.

In the original setup of heterotic superstring theory with six compact dimensions 
\cite{Candelas1985en} it is assumed that the ten-dimensional space-time is the product of a maximally 
symmetric four-dimensional space-time and a compact six-dimensional space, with the property 
that supersymmetry is preserved in the resulting four-dimensional theory. This implies 
that the internal space must be a complex semi-K\"ahler \footnote{Meaning the divergence (co-derivative) of the K\"ahler form vanishes.} manifold with SU(3) holonomy 
and vanishing first Chern class; the exterior derivative of the K\"ahler form being proportional to the 
$H_3$-flux in the internal space. A non-vanishing $H_3$-flux also prevents the internal manifold from being Ricci-flat. 
It has been argued that such an internal space would have a curvature of the order of the string scale, implying
a potential inconsistency with the use of the supergravity approximation. For this reason, only solutions with 
vanishing $H_3$-flux in the internal space given by a Ricci-flat, K\"ahler (Calabi-Yau) manifold  have been considered. 
However, in \cite{Strominger:1986uh}, it was shown that one can turn on $H_3$-fluxes and consider 
a non-Ricci flat internal space, provided one changes the ansatz for the ten-dimensional metric to describe 
a warped compactification, with a warping function proportional to the dilaton. The flux introduces torsion in the 
internal manifold, which is not a K\"ahler manifold, anymore, but can be related conformally to a Calabi-Yau space. 
The study of non-K\"ahler compactifications is further developed in \cite{nonKahler,BPSaction} (see also \cite{Becker:2003dz} 
for an early review). The focus of all these works has been on finding ten-dimensional target space-times
with a maximally symmetric four-dimensional space-time compatible with supersymmetry, and the problem 
of moduli stabilization has not been discussed, except for the overall K\"ahler modulus, which is stabilized by 
the balance between the flux and the exterior derivative of the complex structure. 
All the conditions discussed in \cite{Candelas1985en,Strominger:1986uh} (which are frequent starting points of 
subsequent works) are only valid for ``vacuum solutions,'' and the moduli fields have flat directions.

In applications of string theory to cosmology, we would like to determine the full-fledged effective potential 
of the moduli fields, in order to study the cosmological dynamics it determines. In the literature, studies are usually
limited to searches of minima of the effective potential, since one is only interested in finding stabilization mechanisms. 
For this purpose, one does not have to know the complete shape of the effective potential (see however 
\cite{Cicoli:2013rwa} for a discussion of perturbative and non-perturbative contributions to the potential). 
In this work, we invoke $H_3$-fluxes and gaugino condensation \cite{gaugino} to stabilize 
the axio-dilaton $S$, as in \cite{Gukov:2003cy}. 
Some discussion of dilaton stabilization by the gaugino condensation mechanism (which has been developed for
Calabi-Yau compactifications) can be found in \cite{BPSaction, hetgaugino}. But that discussion concerns primarily
the minima of the effective potential and not its complete shape. After stabilizing $S$, $\text{Re}T$ turns out to be 
a flat direction in field space originating in the ``no-scale structure'' of the K\"ahler potential. In this paper we will
determine leading-order corrections to the effective potential. The result is a runaway potential for the field 
$\text{Re}T$ related to $\sigma$, causing $\sigma$ to slowly evolve towards $\infty$.

In Section \ref{model}, we derive the effective potential in the four-dimensional supergravity action functional 
for the two moduli fields related to the complex scalar field $z=\sigma + i\theta$ introduced above. This is accomplished 
by invoking a standard dimensional reduction procedure applied to the original ten-dimensional superstring 
theory; see also Appendix A. We then describe the gaugino condensation mechanism stabilizing the dilaton. 
Subsequently, we consider corrections to the effective action of the axio-dilaton field arising from quantum
 fluctuations in a non-supersymmetric background. These corrections lead to the 
 effective potential used in our cosmological scenario. Details of the derivation of quantum corrections 
to the effective potential are presented in Appendix B. In Appendix C, we show that $\alpha^{\prime}$ corrections to the 
effective potential are \textit{smaller} than the corrections arising from quantum fluctuations and hence do not play an 
important role in our applications to cosmology.

In Section \ref{parameters}, we discuss conditions on the compactification of the six extra dimensions of string theory 
required in order to obtain a realistic effective potential accounting for both Dark Matter and Dark Energy. While we 
do not  solve the ``coincidence' problem'', i.e., the mystery of why the contribution of Dark Energy to the 
total energy budget of the Universe is becoming important precisely at the current stage of its evolution is left 
unresolved, we show that the fine-tuning of parameters of the string theory required by phenomenology is not 
very severe.

In our analysis we initially assume that all six internal spatial dimensions expand isotropically, as described 
by a single dynamical K\"ahler-modulus field $T$. Under this assumption the resulting potential for the Dark Energy field $\sigma$ is too steep \footnote{We thank Osmin Lacombe for alerting us to this issue.}. We therefore test the assumption that only one of the six internal dimensions is expanding, the other five dimensions being static. The resulting potential of $\sigma$ then turns out to be sufficiently flat to yield an accelerated expansion of the Universe.

\section{Embedding of a Dark Sector model in String Theory} \label{model}

Superstring theory\footnote{see, e.g., the textbook \cite{Pol}} describes degrees of freedom propagating in 
ten-dimensional space-times. An effective field theory on a $(3+1)$-dimensional space-time can be obtained 
from superstring theory by compactifying six of the nine spatial dimensions. The theory thus obtained typically 
contains a number of complex scalar fields that might serve as candidates for the fields $\sigma$ and $\theta$ 
appearing in the dark sector model of \cite{us}. To begin with, there is the model-independent 
axio-dilaton (modulus) field, $S$, which we write as
 \be
 S \, = \, e^{-\Phi} + i a \, ,
 \ee
 where 
 \be
 \Phi \, = \, 2 \phi - {\rm{ln}}(\sqrt{g_6})
 \ee
is the shifted dilaton ($g_6$ is the determinant of the metric of the six-dimensional compactified manifold, $\phi$ 
is the  ten-dimensional dilaton), and $a$ is the axion originating from the universal Kalb-Ramond 2-form field, 
$B_2$, via
 \be \label{KRB}
 da \, = \, e^{-2\Phi} (*_4dB_2) \, .
 \ee
Furthermore, there are moduli fields related to the compactification of the six extra spatial dimensions. 
If all of them expanded isotropically, there would be a single volume-modulus field, 
$T$, given by
 \be \label{vol modulus}
 T \, = \, e^{\Psi} + i b \, ,
 \ee
where $\Psi$ is proportional to ${\rm{ln}} (R)$, with $R^6$ proportional to the volume of the six-dimensional 
compactified internal space, and the field $b$ in \eqref{vol modulus} is another axion field.  

In Eq.~\eqref{vol modulus}, the modulus corresponding to the \textit{overall volume} of the compactified 
internal manifold appears. In principle, the size of the internal manifold could vary independently in different directions. 
This would give rise to further size-moduli fields. There are also shape-moduli fields parametrizing the shape of 
the internal manifold. In the following, we pursue a minimal approach and consider only the two complex
fields $S$ and $T$ introduced above; the reason being that we would like our analysis to be as model-independent 
as possible. Moreover, we (temporarily) neglect the axion field $b$. 
We are thus left with \textit{three real scalar} fields, $\Phi, \Psi$ and $a$. Since we know that after dimensional 
reduction we obtain a supergravity theory, these fields appear in components of superfields and their potential is 
determined by the K\"ahler potential and a holomorphic superpotential $W$ for those superfields using the 
standard formulas (see e.g. \cite{Wess:1992cp} for a textbook treatment).

Next, we recall the expression for the K\"ahler potential, $K$, appearing in the four-dimensional effective 
supergravity action derived from ten-dimensional superstring theory (we use the conventions of \cite{Pol}):
\be
    \kappa^2 K \, = \, -\ln (S+\overline{S}) - 3 \ln (T + \overline{T}) \, ,
\ee
where $\kappa$ is the inverse of the \textit{four-dimensional} Planck mass, $m_{pl}$.

The standard mechanism leading to pinning of the expectation value of the dilaton in heterotic string phenomenology 
employs {\it gaugino condensation} \cite{gaugino}. It yields a contribution to the superpotential of the form
\be \label{supot}
W \, = \, W_0 - Ae^{-a_0S} \, ,
\ee
where $a_0, A$ and $W_0$ are constants. Note that $A$ and $W_0$ have mass dimension three, while $a_0$ 
is dimensionless. The potential $W$ is generated by fluxes on the internal manifold whose integrals are given by 
topological invariants and are therefore \textit{independent} of the volume modulus $T$. The sizes of $W_0$ 
and $A$ are determined by the string scale, $m_s$. As we shall see, in order to realize the required hierarchy 
between the string scale and the Dark Energy scale, $W_0$ has to be suppressed as compared to $A$ 
(see Section \ref{parameters}).

In order to compute the potential resulting from the expressions for $K$ and $W$ above, we need to make 
use of the K\"ahler covariant derivatives of $W$, which are given by
\begin{align}
   D_S W &:= \partial_S W+ W \partial_S ( \kappa^2 K) = \frac{1}{S+\overline{S}}\left[-W_0 + Ae^{-a_0 S}\left(1+a_0(S+\overline{S})\right)\right],\\
    D_T W &:= \partial_T W + W \partial_T (\kappa^2 K) = \frac{3}{T+ \overline{T}}\left[-W_0 + Ae^{-a_0 S}\right],
\end{align}
while the components of the K\"ahler metric are
\begin{equation}
    \kappa^2 K_{S\overline{S}} = \frac{1}{(S+\overline{S})^2}, \quad  \kappa^2 K_{T\overline{T}} = \frac{3}{(T+\overline{T})^2}, \quad K_{S\overline{T}} = 0 \, .
\end{equation}
Using these expressions, one can verify that the potential in the effective supergravity theory is given by
\begin{align}
    V(S,T) &:= e^{\kappa^2 K}\left(K^{I\overline{J}}D_I W\overline{D_J W} - 3 \kappa^2 |W|^2\right)\nonumber\\
    &= \frac{\kappa^2}{S+\overline{S}}\frac{1}{(T+\overline{T})^3}\left \{\left|-W_0 + Ae^{-a_0S}\left(1+a_0(S+\overline{S})\right)\right|^2 +\right. \nonumber\\
    &\left.+3\left| -W_0 + Ae^{-a_0 S} \right|^2 - 3 \left|W_0 - A e^{-a_0 S}\right|^2\right\} \, ,
\end{align}
In terms of the real components $\text{Re} S, a$ and $\text{Re} T$ the potential becomes\footnote{As mentioned 
previously, we are neglecting the axion field $b$. We expect a potential for $b$ to be generated by instanton 
effects. This would imply that $b$ is massive. Furthermore, it is likely that the coupling of $b$ to the Standard Model 
(SM) sector is very weak, since it comes with the modulus field in $T$, which is expected to have only a tiny
coupling to the SM fields. In conclusion, the field $b$ might also contribute to Dark Matter; but we neglect its
contribution henceforth.}
\begin{align}
    V(\text{Re}S, a, \text{Re}T) =&  \frac{\kappa^2}{8(\text{Re}T)^3} \Big[ A^2 \Big( a_0 +
 \frac{1}{2} \text{Re}S^{-1} \Big)^2 \text{Re}S e^{-2a_0\text{Re}S} +
  \frac{W_0^2}{4}\text{Re}S^{-1} \nonumber \\
 & - AW_0 \Big( a_0 + \frac{1}{2}\text{Re}S^{-1} \Big) e^{-a_0 \text{Re}S} \cos(a_0a) \Big] \,. \label{4dpotential}
\end{align}
Expanding $S = e^{-\Phi}+ia$ and $T = e^{\Psi}+ib$ in terms of the canonically normalized fields $\phi$ and $\sigma$, 
defined by
\begin{equation} \label{rescaling}
    \frac{\phi}{m_{pl}} = \frac{1}{\sqrt{2}}\Phi, \quad  \frac{\sigma}{m_{pl}} \equiv \psi = \sqrt{\frac{3}{2}}\Psi,
\end{equation}
the effective action for the fields $\phi$, $\sigma$ and $a= \theta/m_{pl}$ coupled to the four-diimensional metric, but disregarding all other fields, is given by
\begin{equation} \label{Skin}
    S= \int d^4x \sqrt{-g}\left[\frac{1}{2\kappa^2}R -\frac{1}{2}\partial_\mu \phi \partial^\mu \phi - \frac{1}{2}\partial_\mu \sigma \partial^\mu \sigma -\frac{e^{2\sqrt{2}\phi / m_{pl}}}{4} m_{pl}^2 \partial_\mu a\partial^\mu a - V(\phi, \sigma, a)\right],
\end{equation}
where the potential $V$ is given by
\begin{align} \label{pot0}
    V(\phi, \sigma, a) &= \frac{e^{-\sqrt{6}(\sigma/m_{pl})}}{8}e^{-\sqrt{2}(\phi/m_{pl})} \kappa^2 
    \Big[ A^2 e^{-2a_0e^{-\sqrt{2}(\phi/m_{pl})}}\Big(a_0 + \frac{1}{2}e^{\sqrt{2}(\phi/m_{pl})}\Big)^2 +\nonumber \\
    &+ \frac{W_0^2}{4}e^{2\sqrt{2}(\phi/m_{pl})}
    - AW_0\Big(a_0 +\frac{1}{2}e^{\sqrt{2}(\phi/m_{pl})}\Big)e^{\sqrt{2}(\phi/m_{pl})}e^{-a_0e^{-\sqrt{2}(\phi/m_{pl})}}\cos(a_0 a)\Big] \,.
\end{align}
In this expression the fields $\phi$ and $\sigma$ are in four-dimensional Planck units.  

The key feature of this expression for $V$ used in our study is that this potential decreases exponentially 
as a function of $\psi=\sigma/m_{pl}$: 
There is a common factor of $e^{-\sqrt{6} \psi}$ multiplying a non-trivial function of $\phi$ and $a$. A minimum 
of the potential $V$ in the $a$-direction occurs at $a = 0$, and $V$ is quadratic about this minimum. The potential also 
has a minimum in the $\phi$ direction, a feature that guarantees the required stabilization of the dilaton. We denote the value 
of $\phi$ minimizing $V$ at $a = 0$ by $\phi_0$ and the corresponding value of $S$ by $S_0$. At the point 
$a=0, \phi = \phi_0$, the potential $V$ vanishes for all values of $\psi$, and hence, before
correction terms are taken into account, it does not contain any terms that could yield the term proportional to
$\Lambda$ in expression (\ref{toy}).

There are two sources of corrections to this potential that have to be considered: the first one comes from one-loop 
quantum corrections obtained by integrating out quantum fluctuations about the minimum 
of the gaugino potential; and the second one comes from $\alpha^{\prime}$ corrections to the K\"ahler potential. 
These corrections will be discussed in Appendix B and Appendix C, respectively.

Integrating out quantum fluctuations of the axio-dilaton field $S$  yields the following one-loop contribution to the potential 
(see Appendix B):
\begin{equation} \label{pot1}
    V_{1l} \approx \frac{\Lambda_c^2\gamma^2}{16\pi^2}e^{-\sqrt{6}\psi} - \frac{\gamma^4}{64\pi^2}\left[\frac{25}{3}+14(2a_0S_0)^2+\left(2+12(2a_0S_0)^2\right)\left(-\sqrt{6}\psi+\ln\left(\frac{(2a_0S_0)^2\gamma^2}{\bar{\Lambda}^2}\right)\right)\right]e^{-2\sqrt{6}\psi} \, ,
\end{equation}
where, for large values of $\psi$, the first term dominates. In the expression given in \eqref{pot1}, $\gamma$ is related to the gravitino mass,\footnote{Note that there may be an additional contribution to the gravitino mass due to supersymmetry breaking in the matter sector.} $m_{{\tilde{g}}}$, via
\begin{equation} \label{psimass}
    m_{\tilde{g}}^2 = \kappa^4 \left\langle e^{ \kappa^2 K}|W|^2 \right\rangle = \kappa^4 \frac{\left(2Aa_0S_0e^{-a_0S_0}\right)^2}{2S_0(T+\bar{T})^3} = \kappa^4 \frac{A^2a_0(2a_0S_0)e^{-2a_0S_0}}{8}e^{-\sqrt{6}\psi}=: \gamma^2 e^{-\sqrt{6}\psi},
\end{equation}
and $\Lambda_c$ is the energy scale where the four-dimensional effective supergravity analysis breaks down. The value of 
$\Lambda_c$ is proportional to the mass, denoted by $m_{\text{KK}}$, of the Kaluza-Klein states related to the extra dimensions, and thus inversely proportional to the diameter $R_{\rm{KK}}$ of the 6D compact space (see Appendix D for a discussion of this point):
\begin{equation} \label{cutoffscaling}
    \Lambda_c^2 = \frac{1}{R_{\text{KK}}^2} = 
    m_{pl}^2 \left( \frac{m_s}{m_{pl}}\right)^{8/3} e^{\Phi} e^{- \sqrt{2/3} \psi} \, .
\end{equation}
Inserting this expression for $\Lambda_{c}$ in our expression (\ref{pot1}) for the one-loop effective potential, 
we find that the latter scales as $e^{- 4 \sqrt{2/3} \psi}$ (with $\psi= \sigma/m_{pl}$).

In Appendix C we discuss corrections to the potential $V$ originating from $\alpha^{\prime}$ corrections to the K\"ahler potential. 
For large values of $\psi$, the leading-order corrections are given by the $(\alpha^{\prime})^2$- and $(\alpha^{\prime})^3$ 
terms and scale as
\be \label{Kcorr}
\delta V \, \sim \, \kappa^4 \frac{|W|^2}{2 S_0} e^{- 6 (n + 1) \sqrt{2/3} \psi} \, ,
\ee
where $n = 2/3$ for $(\alpha^{\prime})^2$ corrections, and $n = 1$ for $(\alpha^{\prime})^3$ corrections. Moreover, 
the $(\alpha')^2$ corrections are absent in the case of a standard embedding \cite{Anguelova:2010ed}.  Note that 
the right sides of \eqref{psimass} and \eqref{Kcorr} contain the same factor of $\langle e^K|W|^2\rangle$. But, 
for large values of $\psi$, the decay of the K\"ahler correction terms is faster than the decay of the 
leading term in the one-loop quantum corrections considered before. 

We conclude that, for large values of $\psi$, the full-fledged effective potential is well approximated by the sum of the bare 
potential (\ref{pot0}) and the one-loop corrections (\ref{pot1}).  This potential has the same general properties as the potential 
of the dark sector model proposed in \cite{us}, with the volume modulus field $\psi= \sigma/m_{pl}$ as a candidate for the Dark Energy field, 
and the axion $a\, \propto \theta/m_{pl} \,\,(\text{with } f=m_{pl})$ as  the Dark Matter field.

We expect that an effective potential with an exponential dependence on a volume-modulus field $\psi$ and a periodic dependence on an axion field $a$, predicting an axion mass 
that decreases \textit{exponentially} in $\psi$, is a general feature of models emerging 
from superstring theories.

Unfortunately, in the simple scenario considered above, the potential of the field $\sigma$ is too steep to yield Dark Energy. Inserting (\ref{cutoffscaling}) into (\ref{pot1}) we find that
\be
V(\sigma) \, \sim \, e^{- 2 \sigma / f} \, 
\ee
with 
\be \label{exponent1}
\frac{2}{f} \, = \, 4 \sqrt{2/3} m_{pl}^{-1} \, .
\ee
The equation of state parameter $w = p / \rho$, with $p$ being the pressure and $\rho$ the energy density, is given by
\be
w \, = \, - 1 + \frac{1}{3} \left( \frac{2 m_{pl}}{f}  \right)^2 \, ,
\ee
where $w \leq - 1/3$ is required to obtain accelerated expansion. For the exponent given by (\ref{exponent1}) this condition is clearly violated. We must therefore refine our construction.

Similarly to what has been proposed in order to obtain inflation from {\it LVS} (large volume string 
compactification) scenarios \cite{LVS}, we test the assumption that only one of the six internal dimensions of space 
changes in size (extension). The corresponding modulus field is denoted by $T_s \equiv e^{\Psi_s}$, the modulus 
field determining the volume of the remaining five dimensions by $T_0$. This entails three changes. First, the 
exponent in the one-loop potential (the analog of (\ref{pot1})) becomes smaller. Second, the $T$-dependence of 
the cutoff scale $\Lambda_c$ is modified: If the dynamical cycle is short, as compared to the other ones (which 
is assumed in the LVS scenarios) then $\Lambda_c$ is independent of $T$. This further reduces the steepness 
of the potential. To find the exact value of the coefficient in the exponent of the exponential potential of $\sigma$, 
we must address a third change, namely the change in the relation between the field $\Psi$ and the canonically 
normalized field $\psi$.

We first consider the relation between the fields $\Psi$ and  $\psi$. The starting point is the revised 
form of the K\"ahler potential
\be
\kappa^2 K \, = \, - {\rm{ln}}(S + \overline{S}) - \frac{5}{2} {\rm{ln}}(T_0 + \overline{T_0})
- \frac{1}{2} {\rm{ln}}(T_s + \overline{T}_s) \, .
\ee
The part of the kinetic action that depends on $T_s$ then becomes
\be
K_{T_s \overline{T}_s} \partial_{\mu} T_s \partial^{\mu} T_s \, = \, \frac{1}{8}  \partial_{\mu} \Psi_s \partial^{\mu} \Psi_s \, ,
\ee
wherefrom it follows that the canonically normalized field $\psi$ derived from $\Psi_s$ satisfies
\be
\Psi_s \, = \, 2 \psi \, .
\ee
We note that for $n$ dynamical internal dimensions, all of the same radius, the relation would be 
$\Psi = \frac{2}{\sqrt{n}} \psi$.

As observed at the end of Appendix B, results similar to (\ref{psimass}) and (\ref{pot1}) are obtained 
in a scenario involving anisotropic extra dimensions, namely
\be
V_{1l} \, \sim \, \Lambda_c^2 e^{\kappa^2 K} \, .
\ee
The key point is that $e^K$ now scales as $e^{\Psi_s / 2}$ instead of as $e^{3 \Psi}$. If $\Lambda_c$ is independent 
of $\psi$ (which is the case if $T$ does not correspond to the largest one of the internal spatial dimensions), then we 
obtain that, in terms of the canonically normalized field $\psi$,
\be
V_{1l} \, \sim \, e^{\psi} \, ,
\ee
which is sufficiently flat to yield Dark Energy: The resulting equation of state parameter $w$ has the value
$w = - 2/3$, which is still slightly too large to be compatible with current observational constraints obtained 
in the context of the standard cosmological model. But our findings provide a ``proof of concept'' that
Dark Energy could result from heterotic superstring theory.

\section{Constraints on Parameter Values} \label{parameters}

In the previous section we have shown that a potential of the form given in \eqref{toy} arises quite naturally from 
superstring theory. The key question is then whether, without excessive fine-tuning, one obtains coefficients 
multiplying the different terms in the potential \eqref{toy} that are compatible with phenomenology.
In order for the axion-independent term in the potential (the first term in (\ref{pot1})) to have the right order of 
magnitude to describe the Dark Energy density observed today, we must have that \footnote{Here we use the exponents corresponding to an isotropic time-dependent internal space. The only change when passing to the anisotropic case concerns the value of the coefficient in the exponent, which is not relevant here.}
\be \label{cond1}
\frac{\Lambda_c^2 \gamma^2}{16 \pi^2} e^{-3 \sqrt{2/3} \psi} \, \sim \, T_o^4 z_{eq} \, ,
\ee
where $T_0$ is the current temperature of the cosmic microwave background, and $z_{eq} \sim 10^4$ is the redshift corresponding to the time of equal matter and radiation. The right hand side of this equation is the current matter energy density. 

Inserting the value of the cutoff scale $\Lambda_c$ from (\ref{cutoffscaling}) and the expression for $\gamma$ from (\ref{psimass}), the condition (\ref{cond1}) becomes
\be \label{criterion}
\frac{1}{64 \pi^2} \frac{A^2}{m_{pl}^2} \left( \frac{m_s}{m_{pl}}\right)^{8/3} a_0^2 
e^{-2 a_0 S_0} e^{-4 \sqrt{2/3} \psi} \, \sim \, T_0^4 z_{eq} \, .
 \ee
To further evaluate this condition we use the fact that \cite{Pol}
\be
A \, \sim \, C_H m_s^3 \, ,
\ee
where $C_H$ is the Coxeter number of the gauge group whose gaugino is condensing.  The number $C_H$ also yields the order of magnitude of the constant $a_0$  appearing in the superpotential \cite{Gukov:2003cy} 
\be \label{a0value}
a_0 \, \sim \, \frac{8 \pi^2}{C_H} \, .
\ee
Inserting these expressions in (\ref{criterion}), we obtain the following condition on the combination of the dilaton expectation value, given by $S_0$, and the present value of the Dark Energy field $\psi$:
\be \label{cond2}
e^{- 2 a_0 S_0} e^{- 4 \sqrt{2/3} \psi} \, \sim \, 
\frac{1}{2 \pi^2} \left( \frac{m_{pl}}{m_s}\right)^{6 + 8/3} \left(\frac{T_0}{m_{pl}}\right)^4 z_{eq} \, .
\ee
Working with the string scale $m_s \sim 10^{16} {\rm{GeV}}$ this gives
\be \label{expcond}
e^{- 2 a_0 S_0} e^{- 4 \sqrt{2/3} \psi} \, \sim \, 10^{-97} \, ,
\ee
or, equivalently, 
\be \label{cond3}
2 a_0 S_0 + 4 \sqrt{2/3} \psi \, \sim \, 210 \, .
\ee

The bottom line is that it is not implausible that condition (\ref{cond2}) is satisfied, in spite of the large hierarchy between 
the string scale and the scale of Dark Energy (given by $T_0$ above). We can either assume that the current value of 
$\psi $ is significantly larger than $1$, or else we may assume that the value of the stabilized dilaton is small 
(i.e., the value of $S_0$ is large). Since large values of $S_0$ and of $\psi$ lead to exponential suppression of the 
left side of (\ref{cond2}), the tuning of parameters required in order for \eqref{cond2} to hold is not very severe. 
For example, if we assume that $\psi$ is not larger than $1$ then a value of $a_0 S_0 \sim 10^2$ is required. 
The amount of tuning of $S_0$ required here depends on the value of $a_0$ and hence on $C_H$. The least amount of tuning necessary to satisfy the 
phenomenological constraints is obtained for the smallest possible realistic value of $C_H$, namely $5$ for $SU(5)$. With the resulting value for $a_0$, namely $a_0 \sim 10$, we must require that $S_0 \sim 10$. The value of $S_0$ has consequences for the particle physics phenomenology resulting from the model, because 
\be
S_0 \, = \, \frac{1}{g_{YM}^2} \, ,
\ee
see \cite{Pol}). With $S_0\sim 10$, this relation yields quite a realistic value for the Yang-Mills coupling constant $g_{YM}$ at the scale of Grand Unification.

In order to understand how a large value of $S_0$ can be obtained from string theory, we recall that $S_0$ is determined by minimizing the potential (\ref{pot0}) in the dilaton direction, namely
\be \label{W0rel}
W_0 - A e^{- a_0 S_0} (1 + 2 a_0 S_0) \, = \, 0 \, .
\ee
Note that $W_0$ is given by the fluxes through the internal manifold \cite{Pol}
\be
W_0 \, \sim \, h m_s^3 \, ,
\ee
where $h$ is the flux in string units. Hence, in order to obtain $(1 + 2 a_0 S_0) e^{- 2 a_0 S_0} \ll 1$ we require a very small flux $h \ll 1$.

Thus, while our model does not predict the value of the currently observed value of the Dark Energy density, 
our analysis shows that it is possible to obtain a realistic value from string scale physics without very severe 
fine-tuning of parameters. This may come as a surprise.

The axion mass, $m_{\theta}$, can be read off from (\ref{pot0}). Taking into account that the canonically normalized 
axion field $\theta$ is given in terms of the dimensionless field $a$ by 
\be
\theta \, = \, \frac{e^{\Phi}}{\sqrt{2}} \, m_{pl}\, a \, ,
\ee
see (\ref{Skin}), making use of (\ref{W0rel}) to replace $W_0$ by $A$ and $S_0$, and inserting the expressions for 
$a_0$ and $A$ yields
\be
m_{\theta}^2 \, \simeq \, (8 \pi^2)^4 C_H^{-2} S_0^3  
 \left( \frac{m_s}{m_{pl}}\right)^6 m_{pl}^2 e^{-2 a_0 S_0} e^{-3 \sqrt{2/3} \psi} \, .
\ee 
Inserting condition (\ref{expcond}), required to obtain the observed hierarchy between the string scale and the Dark Energy scale, then yields
\be
m_{\theta}^2 \, \simeq \, (8 \pi^2)^4 C_H^{-2} S_0^3  
 \left( \frac{m_s}{m_{pl}}\right)^6 m_{pl}^2 10^{-97}  e^{\sqrt{2/3} \psi} \, .
\ee 
With the numbers for $S_0$ and $C_H$ used in the above discussion of the amplitude of the $\psi$ field potential, 
we obtain
\be
m_{\theta} \, \sim 10^{-24} e^{\sqrt{1/6} \psi}{\rm{eV}} \, ,
\ee
and thus our axion is an ultralight boson. In order to be consistent with the lower bound on the mass of an ultralight axion 
which is of the order of $10^{-22} {\rm eV}$ (see e.g. \cite{ALPrevs} for recent reviews), we require a value of $\psi$ 
at the beginning of the Dark Energy phase that is slightly larger than $1$, which is consistent with the desired phenomenology (see \cite{us}).

Before considering contributions from other sectors of string theory (e.g., the sector containing the degrees 
of freedom of the Standard Model), the gravitino mass of our model is given by (\ref{psimass}) which is of the 
order $10^{-37} {\rm eV}$. At first sight, this seems to be in conflict with the lower bound on the gravitino mass 
in \cite{PDB}, which is based on missing transverse momentum studies at the LHC \cite{Atlas}. Note, however, 
that this bound holds for one specific class of supersymmetry-breaking scenarios, namely gauge-mediated 
supersymmetry breaking \footnote{When talking about ``spontaneous breaking of local supersymmetry'' one must remember that the action is still invariant under supersymmetry, and there should be a gauge-invariant formulation of the super-Higgs mechanism, similar to the one reviewed in \cite{Maas:2017wzi} (see \cite{Frohlich:1980gj} for original references) for the usual Higgs mechanism.}, which leads to direct couplings of the gravitino, squark and gaugino to the Standard Model 
gauge sector. Thus, as long as we consider a mechanism of supersymmetry breaking that does not contain such vertices, 
the bound in \cite{Atlas} does not apply (see, e.g., \cite{Wei} for such a mechanism, which was proposed in the context of 
String Gas Cosmology \cite{BV, SGCrev}). Furthermore, as discussed in the next paragraph, there must be a source 
of supersymmetry breaking unrelated to the gravitational sector of string theory we are considering here. This could yield 
further contributions to the gravitino mass.
 
Another question to be addressed is how a tiny value of the gravitino mass, such as the one obtained above, 
may be compatible with lower bounds on the masses of the superpartners of the Standard Model fields. 
Since our analysis is limited to the gravitational sector of a supergravity model emerging from 
string theory, the Standard Model fields do not appear in it. In string phenomenology involving a Minkowski 
spacetime background, a supersymmetry breaking mechanism must be introduced, and the energy scale 
of the resulting symmetry breaking is usually expected to be high (because the string scale is the 
characteristic energy scale in the theory).  In our model, we have, in addition to 
supersymmetry breaking caused by the non-trivial cosmological background, a symmetry breaking mechanism 
inherent in the Standard Model sector. Hence, if this mechanism does not generate sizable contributions to 
the effective action of the gravitational sector problems in reconciling the large energy scale of the superpartners 
of the Standard Model fields with the tiny gravitino mass of our model will not arise.  
A similar argument has recently been brought forward in \cite{Cliff} (see also \cite{Cliff2}), where non-linearly realized supersymmetry \cite{nonlinear} is invoked to couple non-supersymmetric matter to a supersymmetric gravitational sector.

\section{Discussion and Conclusions} \label{conclusion}

In this paper we have described a rather compelling derivation of the unified dark-sector model proposed 
in \cite{us} from superstring theory: The real part of a universal superfield $T$, which is related to 
the  variable size of one of the six dimensions of the compactified internal space, 
yields the Dark Energy field, while the axion of the universal axio-dilaton superfield $S$ gives rise to 
Dark Matter. The combination of gaugino condensation with one-loop effects of quantum fluctuations 
in a non-supersymmetric cosmological background yields an effective potential of a form resembling 
the one postulated in \cite{us}. In our scenario, the Dark Energy field behaves like a standard quintessence 
field that rolls down an exponentially decreasing potential, while the Dark Matter field describes an ultralight axion.

In this paper we do not solve the ``coincidence problem,'' namely the fact that Dark Energy is becoming dominant 
precisely at the present time. But we have shown that the observed magnitude of the Dark Energy density can be obtained 
from a string-theory construction without very severe fine-tuning. Demanding that the dilaton be stabilized at a sufficiently 
small value, as expressed in (\ref{criterion}), is all that is required; and this can be obtained by some tuning of 
the flux through the internal manifold.

It would be good to identify the origin of those features of the effective potential that cause the 
runaway direction necessary to model dynamical Dark Energy. This would help understanding whether and how a suitable 
effective potential can be constructed in type IIB string theory. First, it is clear that the vanishing of the 
tree-level potential, after dilaton stabilization, is a generic feature of the ``no-scale'' K\"ahler potential 
for the volume modulus, resulting in a four-dimensional space-time incompatible with
supersymmetry for any value of $\text{Re}T$. Second, the presence of perturbative $\alpha'$- and quantum 
corrections for a non-supersymmetric background space-time is a general feature, and if we do not tune
parameters in such a way that these contributions will compete among themselves then the resulting 
potential for the volume modulus has a runaway direction. Finally, regardless of which correction dominates, 
the resulting runaway scale depends on the value of the superpotential evaluated at the background solution. 
For heterotic string theory, we have been able to show that this scale can be tuned in such a way
that the volume modulus plays the role of the quintessence field. Focusing on the gravitational sector, 
we expect that all these features are also present in type IIB string theory, even though the dilaton stabilization mechanism 
due to $G_3$-fluxes is different from the gaugino mechanism used in heterotic string theory.

\vspace{0.4cm}

\noindent \textbf{Note added:} After completing our draft, two papers appeared \cite{Cicoli:2021fsd} where the problem
of obtaining Quintessence from string theory moduli fields is discussed, and where it is argued that it is not possible 
to obtain a slow-roll solution from the runaway behaviour of the potential for $S$ and $T$ obtained in the limit 
of the moduli space of type IIB string theory. In the present paper we go beyond the regime considered in those
papers, since, after stabilizing the axio-dilaton field, we consider one-loop quantum corrections in the resulting 
background space-time, starting from heterotic string theory.

\section*{Acknowledgement}
\noindent  RB thanks the  Pauli Center and the Institutes of Theoretical Physics and of Particle Physics and 
Astrophysics of the ETH for hospitality. The research at McGill is supported in part by funds from NSERC 
and from the Canada Research Chair program. HB would like to thank Keshav Dasgupta for discussions and 
the Nordic Institute for Theoretical Physics for the hospitality during the late stage of this work. HB's research 
is supported by the \textit{Fonds de Recherche du Qu\'ebec} through the PBEEE merit program, file number 303549. 
We are grateful to Osmin Lacombe for alerting us to the issue of the steepness of the potential derived 
in the initial version of this paper.

\section*{Appendix A: Review of the Dimensional Reduction Procedure}

In this appendix we review how, after dimensional reduction, the four-dimensional effective action of Section \ref{model} 
arises from the ten-dimensional effective action predicted by heterotic string theory. We let $D$ and $d$ 
be the dimensions of the higher-dimensional and the lower-dimensional theories, respectively, and we 
set $D=10$ and $d=4$ only at the end of our calculations.

Consider the bosonic part of the string-frame supergravity action derived from heterotic string theory \cite{Pol}
\begin{equation} \label{sugra}
    S = \frac{1}{2\kappa^2_{10}} \int d^D x \sqrt{-G}e^{-2\phi}\left[R + 4\partial_\mu \phi \partial^\mu \phi -\frac{1}{2}|\tilde{H}_3|^2 - \frac{\kappa^2_{10}}{30g^2}\text{tr}|F_2|^2\right] \, ,
\end{equation}
where $\phi$ is the ten-dimensional dilaton, $G$ is the determinant of the ten-dimensional metric, $R$ is its Ricci scalar, 
and $\kappa_{10}$ is the inverse of the ten-dimensional Planck length. At tree level, the massless modes of free heterotic
string theory are related to the the metric, the dilaton and the two form $B_2$, with $H_3 = dB_2$, and a gauge field $A_1$ in the adjoint representation of $SO(32)$ or $E_8 \times E_8$.
 
We wish to dimensionally reduce this theory to $d<D$ dimensions. First, we choose to work in the $D$-dimensional 
Einstein frame, using the Weyl rescaling  
\begin{equation}
    G_{ab} = \Omega^2G_{ab}^{(E)}= e^{\phi/2}G_{ab}^{(E)}.
\end{equation}
This yields
\begin{align}
    \sqrt{-G}e^{-2\phi} R &=\sqrt{-G^{(E)}}e^{(D-10)\phi/4}\left[R^{(E)} - \frac{(D-1)}{2}\nabla_E^2\phi-\frac{(D-2)(D-1)}{16}
    G^{(E)ab}\partial_a \phi \,\partial_b \phi\right]\\
    &= \sqrt{-G^{(E)}}\left(R^{(E)}-\frac{9}{2}\nabla_E^2\phi- \frac{9}{2}G^{(E)ab}\partial_a \phi \,\partial_b \phi\right),
\end{align}
where the superscripts $(E)$ indicate that quantities are taken in the Einstein frame. In the second line we have set $D=10$.  For a $p$-form field strength $F_p$, we use the identities
\begin{equation}
    \int F_p \wedge \ast F_p = \int d^Dx\sqrt{-G}|F_p|^2, \quad |F_p|^2 =\frac{1}{p!}G^{a_1b_1}\cdots G^{a_p b_p}F_{a_1\cdots a_p} F_{b_1 \cdots b_p},
\end{equation}
and
\begin{equation}
    \sqrt{-G}e^{-2\phi}|F_p|^2 = e^{(D-2p-8)\phi/4}\sqrt{-G^{(E)}} |F_p|_{(E)}^2 = e^{-(p-1)\phi/2}\sqrt{-G^{(E)}}|F_p|^2_{(E)}\,,
\end{equation}
where the inner product $|F_p|^2_{(E)}$ has contractions involving the Einstein frame metric.

Applying this result to the heterotic action, we obtain
\begin{equation}\label{Eframe10dheteroticaction}
    S = \frac{1}{2\kappa^2_{10}}\int d^{10}x\sqrt{-G^{(E)}}\left[R^{(E)}-\frac{1}{2}\partial_\mu\phi\partial^\mu\phi -\frac{e^{-\phi}}{2}|\tilde{H}_3|^2_{(E)} - \frac{e^{-\phi/2}\kappa^2_{10}}{30 g^2}\text{tr}|F_2|^2_{(E)}\right].
 \end{equation}

We dimensionally reduce this action using the following metric (we are following the notation of \cite{Gukov:2003cy}, as reviewed in \cite{Baumann})
\begin{equation}
    ds^2_{10} = G^{(E)}_{ab} dx^a dx^b =  e^{-6\sigma(x)}g_{\mu\nu}(x)dx^ \mu dx^\nu + e^{2\sigma(x)} h_{mn}(y)dy^m dy^n,
\end{equation}
where $g_{\mu\nu}$ is the four-dimensional metric, $\sigma$ describes the overall size of the internal manifold and 
$h_{mn}$ is a internal fiducial metric which we take to be time-independent. Note that, with this ansatz, we are 
freezing out all the K\"ahler and complex structure moduli, except for the overall volume modulus of the 
internal space. 

The ten-dimensional Ricci scalar reduces to
\begin{align}
    R^{(E)} &= e^{6\sigma}\left[R(g) + e^{-4\sigma}R(h) + (6(D-1)-8d)\nabla^2\sigma +\right.\nonumber\\
    &\left.+\left(6(D-1)\frac{8d}{2}-\frac{ 8^2d}{4}(d+1)-9(D-2)(D-1)\right)\partial_\mu \sigma \partial^\mu \sigma \right], 
\end{align}
so that
\begin{equation}
    \sqrt{-G^{(E)}}R^{(E)} = \sqrt{-g}\sqrt{h}\left[R(g) + e^{-4\sigma}R(h)+ 6\nabla^2 \sigma - 24 \partial_\mu\sigma \partial^\mu \sigma \right].
\end{equation}
Then, the gravity-scalar part of the four-dimensional action reads
\begin{equation}\label{gravity-scalaraction}
    S = \frac{1}{2\kappa^2_{10}} \int d^4x \sqrt{-g}\tilde{V}_6\left[R(g) + e^{-4\sigma}\langle R(h)\rangle -24\partial_\mu \sigma \partial^\mu \sigma -\frac{1}{2}\partial_\mu \phi \partial^\mu \phi + \cdots\right],
\end{equation}
where $\tilde{V}_6$ is the fiducial internal volume and $\langle R(h)\rangle$ is the mean curvature of the internal metric. The
dots stand for terms involving the axion and the fluxes.

The four-dimensional axion, $a(x)$, which emerges from the space-time components of $H_{3}$, is defined as follows.
\begin{equation}\label{axiondef}
    (\ast_E d a)_{\mu\nu\rho} = e^{-2\Phi}(H_3)_{\mu\nu\rho},
\end{equation}
where $\ast_E$ is the ten-dimensional Hodge dual in the Einstein frame. This definition with the $\Phi$-dependent factor is motivated by the following considerations: In the ten-dimensional Einstein frame, the $H_3$-kinetic term comes from
\begin{equation}
    \int d^{10}x \sqrt{-G^{(E)}} e^{-\phi}|\tilde{H}_3|^2_{(E)} = \int e^{-\phi} \tilde{H}_3\wedge \ast_E \tilde{H}_3,
\end{equation}
where $\tilde{H}_3$ is given by
\begin{equation}
    \tilde{H}_3 = H_3 -\frac{\kappa^2_{10}}{g^2_{10}}\Omega_3,
\end{equation}
with $H_3 =dB_2$ the field strength of the Kalb-Ramond 2-form, and
\begin{equation}
    \Omega_3 = \Omega_3(A) - \Omega_3(w)
\end{equation}
the difference between the gauge-field- and the spin-connection Chern-Simons 3-forms; (we are using the conventions of \cite{Pol}). We have that
\begin{equation}\label{Kalb-Ramond}
    \tilde{H}_3\wedge \ast_E \tilde{H}_3 = H_3\wedge \ast_E H_3 - 2\frac{\kappa^2_{10}}{g^2_{10}}\Omega_3\wedge \ast_E H_3 + \frac{\kappa^4_{10}}{g^4_{10}}\Omega_3\wedge \ast_E \Omega_3\,.
\end{equation}
In terms of the axion field $a(x)$, the first term on the right side is given by
\begin{equation}
    H_3\wedge \ast_E H_3 = e^{4\Phi}\ast_E da \wedge da,
\end{equation}
which, after dimensional reduction, gives rise to the axion's kinetic term: 
\begin{align}\label{axionkinetic}
    \int e^{-\phi}H_3\wedge \ast_E H_3 &= \tilde{V}_6\int e^{-\phi + 12\sigma}e^{4\Phi}da\wedge \ast_4 da\nonumber\\
    &= \tilde{V}_6\int e^{2\Phi}da \wedge \ast_4 da\nonumber\\
    &= \tilde{V}_6\int d^4x \sqrt{-g} e^{2\Phi} g^{\mu\nu}\partial_\mu a \partial_\nu a,
\end{align}
where $\ast_4$ is the four-dimensional Hodge dual. Dimensional reduction of the second term on the right side of
\eqref{Kalb-Ramond}
\begin{align}
    \int e^{-\phi}\Omega_3 \wedge \ast_E H_3 &= \tilde{V}_6 \int e^{-2\Phi} \Omega_3\wedge e^{2\Phi} da\\
    &= \tilde{V}_6 \int \Omega_3 \wedge da \\
    &= \tilde{V}_6 \int a d\Omega_3,
\end{align}
where a total derivative term has been neglected. From the definition of the Chern-Simons 3-form for the gauge field,
\begin{equation}
    \Omega_3(A) = \text{tr}\left(A\wedge dA + \frac{2}{3}A\wedge A\wedge A \right),
\end{equation}
we infer that
\begin{equation}
    d \Omega_3(A) =  \text{tr}\left(F_2\wedge F_2\right), 
\end{equation}
where $F_2$ is the field strength 2-form of the gauge field. Thus, the four-dimensional action has a contribution
\begin{equation}
    \int d^4x \sqrt{-g} \;a(x) \text{tr}(F_2\wedge F_2)\,,
\end{equation}
which explains why there is the $\phi$-dependet factor in the definition of the axion \eqref{axiondef}.

Defining the four-dimensional dilaton, $\Phi$, and the volume modulus, $\Psi$, as (see \cite{Gukov:2003cy})
\begin{equation}
    \Phi = \frac{\phi}{2} - 6\sigma, \quad \Psi = \frac{\phi}{2} + 2\sigma,
\end{equation}
it is straightforward to show that
\begin{equation}
    -\frac{1}{2}\partial_\mu \phi \partial^\mu \phi -\frac{3}{2}\partial_\mu\Psi \partial^\mu \Psi = -\frac{1}{2}\partial_\mu \phi \partial^\mu \phi - 24 \partial_\mu \sigma \partial^\mu \sigma,
\end{equation}
and hence (considering a Ricci flat internal manifold, as in a Calabi-Yau compactification) we see that 
Eq.~\eqref{gravity-scalaraction} becomes 
\begin{equation}\label{4dgravityscalaraction}
    S = \int d^4x \sqrt{-g}\left[\frac{1}{2\kappa^2}R -\frac{1}{4 \kappa^2}(\partial \Phi)^2 -\frac{3}{4 \kappa^2}(\partial \Psi)^2 - \frac{1}{4 \kappa^2} e^{2\Phi} (\partial a)^2+\cdots \right],
\end{equation}
where the four-dimensional gravitational coupling constant is given by $\kappa^2 = \kappa^2_{10}/\tilde{V}_6$. 
The axion's kinetic term is included, as calculated in \eqref{axionkinetic}. The dots stand for terms representing 
the contributions to the potential that come from fluxes and gaugino condensation.

The action in \eqref{4dgravityscalaraction} can be embeded in the bosonic part of the $\mathcal{N} =1$ supergravity action
\begin{equation}
    S = \int d^4x \sqrt{-g}\left[\frac{1}{2\kappa^2}R - K_{I\Bar{J}}\partial_\mu A^I \partial^\mu \Bar{A}^J \right],
\end{equation}
where the K\"ahler metric $K_{I\Bar{J}} = \partial_{I}\partial_{\Bar{J}}K$ is derived from the potential 
\begin{equation}\label{Kahlerpotential}
     \kappa^2 K = - \ln (S+ \Bar{S}) - 3\ln (T+\Bar{T}),
\end{equation}
with $S = e^{-\Phi} + ia$ and $\text{Re}(T) = e^{\Psi}$. (This reproduces the conventions used in \cite{Gukov:2003cy}.)

\section*{Appendix B: Volume potential from quantum fluctuations of the axio-dilaton system}

Here we study the generation of a potential for the volume modulus $T$ by integrating out quantum fluctuations of the axio-dilaton field $S$ around the minimum of the gaugino condensation potential, which is given in (\ref{4dpotential}).

\subsection{Preliminaries}

We begin by recalling the standard calculation of the effective potential in $d$ space-time dimensions. To determine the effective action   we use the background field method. For a single scalar field $\phi$, the (one-loop) Euclidean effective action can be obtained by expanding the functional $S$ about the minimum of the potential, which yields
\begin{align}
    S_{\text{eff}}&= S_0 - \frac{1}{2}\ln \det \frac{\delta^2S_0}{\delta\phi \delta \phi}\\
    &= S_0 -\frac{1}{2}\text{tr}\ln \mathcal{D},
\end{align}
where $S_0$ is the action evaluated at the field configuration $\phi_c$ about which we are expanding, 
and in the last line we have introduced the functional operator $\mathcal{D}: = \delta^2 S_0/\delta \phi \delta \phi$. 
Calculating in momentum space and using an integral representation for the logarithm, we can write the functional trace as
\begin{equation}
    \text{tr} \ln \mathcal{D} = \int_0^\infty \frac{ds}{s}\int \frac{d^dp}{(2\pi)^d}e^{-s\mathcal{D}(p)}.
\end{equation}

If $\mathcal{D}(p) = p^2+w^2$, the momentum integral can be evaluated easily, and we find that
\begin{equation}
    \text{tr}\ln \mathcal{D} = \int_0^\infty\frac{ds}{s(4\pi s)^{d/2}}e^{-w^2 s}.
\end{equation}
Note that $w^2$ is the non-kinetic part of the (quadratic) $\mathcal{D}$ operator. For a general potential $V(\phi)$, 
we have that $w^2 = V''(\phi_c)$.  Hence
\begin{align}
    S_{\text{eff}} &= S_0 -\frac{1}{2}\int_0^\infty\frac{ds}{s(4\pi s)^{d/2}}e^{-w^2 s}\nonumber\\
    &= S_0 -\frac{1}{2}\frac{\Gamma(-d/2)}{(4\pi)^{d/2}}(w^2)^{d/2}.
\end{align}
This result is obviously divergent for $d = 4$; but we can use dimensional regularization to obtain the result for 
$d = 4-2\epsilon$:
\begin{equation}
    S_{\text{eff}} = S_0 - \frac{1}{4}\frac{w^4}{(4\pi)^2}\left(\frac{2}{\epsilon}-\gamma_E-\ln \frac{w^2}{4\pi \mu^2}+\frac{3}{2}\right),
\end{equation}    
where $\mu$ is an (arbitrary) energy scale introduced to get a dimensionless argument in the logarithm. The divergence 
as $\epsilon \to 0$ can be canceled by a proper choice of counterterms in the action, while the finite part yields the 
expression for the effective potential. The running of the coupling constants in $V$ renders the final expression for
the effective potential independent of $\mu$. 
Assuming the $\overline{\text{MS}}$ scheme, we find
\begin{equation}
    V_{\text{eff}} = V_0 + \frac{w^4}{64\pi^2}\ln \frac{w^2}{4\pi \mu^2}.
\end{equation}
The final result depends on the initial potential through $w^2 = V_0''(\phi_c)$. 

Except for the renormalization step, the procedure outlined above can be applied to our model. Note, however, 
that when gravitational interactions are involved, we are not allowed to freely subtract constant terms, and we 
have to impose an ultraviolet cutoff at a scale $\Lambda_c$, above which the effective field theory description 
breaks down.

\subsection{One Loop Corrections in ${\cal{N}} = 1$ Supergravity}

In this subsection we set up the calculation of the corrections to the potential around a supersymmetry breaking configuration in supergravity, for a generic K\"ahler potential and superpotential.

The first fact to notice is that the potential of the model with
\begin{equation}
     \kappa^2 K = -\ln(S+\bar{S})-3\ln(T+\bar{T}), \quad W = W_0 - A e^{-a_0 S},
\end{equation}
has a minimum that breaks supersymmetry, because $D_T W \neq 0$, while the supersymmetry transformation of the fermion in the multiplet of $T$ includes a term proportional to ${D_{\bar{T}} W}$. The supersymmetry breaking scale is then proportional to $D_T W$. In the following we will be using Planck units, i.e. setting $\kappa = 1$.

The one-loop quantum correction of a potential in a general quantum field theory containing fields with arbitrary spins can be read from the effective action:
\begin{equation}
    W_{eff} = -\frac{1}{2}\sum_j(-1)^{2j}n_j\text{tr} \ln (\square + m_j^2) = -\text{Str}\ln(\square + M^2),
\end{equation}
where $n_j$ is the number of propagating degrees of freedom for a spin-$j$ field,\footnote{In four dimensions, massive fields with $j> 0$ have $n_j = 2j+1$ while $n_j = 2$ for massless fields;  $n_j=1$ for a real massless scalar field. This counting only works if we are dealing with irreducible representations of the Lorentz group.} the supertrace takes into account the minus sign in contributions from half-integer spin fields, and the mass matrix, $M^2$, is the second variation of the tree-level action evaluated at a background configuration. 

In order to evaluate the supertrace above, we consider the contribution of each individual spin term:
\begin{equation}
    \text{tr}'\ln(\square + m_j^2)= \int \frac{d^dp}{(2\pi)^d}\ln(p^2+m_j^2) = -\int \frac{d^dp}{(2\pi)^d}\int_0^\infty \frac{ds}{s}e^{-s(p^2+m^2_j)}\,,
\end{equation}
where we are using a momentum-space representation for the trace in the first equality and an integral representation 
for the logarithm in the second step. Usually, we are varying the tree-level action around a static configuration, 
in which case all the fields are minimized and have some fixed mass. In our case, we will use an \textit{adiabatic 
approximation}, in which we neglect the time evolution of the K\"ahler moduli field (which occurs on a cosmic time scale 
much larger than the time scale of the fluctuations); it is fixed at an arbitrary value. Thus, within this adiabatic approximation, 
the mass term shown above is constant in time, and hence we can perform the momentum-space integration 
without any difiiculties:
\begin{equation}
    \text{tr}'\ln(\square + m_j^2)= -\int_0^\infty ds\frac{e^{-sm^2_j}}{s(4\pi s)^{d/2}} = \left(\frac{m_j^2}{4\pi}\right)^{d/2}\Gamma(-d/2).
\end{equation}
In four dimensions, this integral is divergent at the lower limit. We cope with this problem by introducing a UV cut-off 
scale $\Lambda_c$,
\begin{equation}
    \text{tr}'\ln(\square + m_j^2)= \left(\frac{m^2}{4\pi}\right)^{2}\int_{\frac{m_j^2}{2\Lambda_c^2}}^\infty dt\; t^{-3}e^{-t},
\end{equation}
where we change variables to $t = m^2_j s$ in order to obtain a dimensionless integral. Integrating by parts, we get
\begin{equation}
    \text{tr}'\ln(\square + m_j^2)= \frac{1}{8\pi^2}\left(\Lambda_c^4 -\frac{m_j^2\Lambda_c^2}{2}-\frac{m_j^4}{4}\ln\left(\frac{m^2_j}{2\Lambda_c^2}\right)+\frac{m^4_j}{4}\gamma_\text{E}\right).
\end{equation}
The effective potential then becomes
\begin{align}\label{1loopeffectivepot}
   V_{1l} &= \frac{1}{16\pi^2}\sum_j (-1)^{2j}n_j\left(\Lambda_c^4 -\frac{m_j^2\Lambda_c^2}{2}-\frac{m_j^4}{4}\ln\left(\frac{m^2_j}{2e^{\gamma_\text{E}}\Lambda_c^2}\right)\right)\nonumber\\
    &= \frac{1}{16\pi^2} \left(\Lambda_c^4\text{Str}\;\mathbf{1}- \frac{\Lambda_c^2}{2}\text{Str}M^2 - \frac{1}{4}\text{Str}M^4\ln\left(\frac{M^2}{\bar{\Lambda}^2}\right)\right) \, ,
\end{align}
where we have defined ${\bar{\Lambda}^2} = 2e^{\gamma_E} \Lambda_c^2$. In the following we apply this result to our model.

We are considering a theory with two chiral superfields, one corresponding to the axio-dilaton field and the other one 
to the K\"ahler modulus field. The complete action for the system can be found in \cite{Wess:1992cp}. On top of the 
bosonic sector already described, it contains the kinetic terms for the spin-1/2 partners of $S$ and $T$, the kinetic 
term for the gravitino, four-fermion interaction terms and Yukawa-like couplings between the scalar fields and the
fermions fields. Around a given bosonic background, the Yukawa-like terms become mass terms for the fermions, 
yielding a non-trivial fermionic mass matrix. The bosonic mass matrix can be inferred from the variations of the 
four-dimensional potential. These matrices can be written compactly in terms of a generalized potential $G$, defined by
\begin{equation}
    G = K + \ln W + \ln \bar{W}.
\end{equation}
The gravitino mass is then given by 
\begin{equation}
    m_{\tilde{g}} = \langle e^{G/2} \rangle,
\end{equation}
while the spin-1/2 mass matrix is
\begin{equation}
    (M_{1/2})_{ij} = \left\langle \left(\nabla_i G_j + \frac{1}{3}G_i G_j\right)e^{G/2}\right\rangle ,
\end{equation}
and the spin-0 mass matrix is
\begin{equation}\label{bosonicM2}
    (M_0)^2 = \begin{pmatrix} 
    M^2_{i\bar{j}} & M^2_{ij} \\
    M^2_{\bar{i}\bar{j}} & M^2_{\bar{i}j} \end{pmatrix},
\end{equation}
and can be obtained from variations of the potential as reviewed below.

Starting from the potential
\begin{equation}
    V= e^{G}\left(g^{i\bar{j}}G_iG_{\bar{j}}-3\right),
\end{equation}
the variations are found to be
\begin{align}
    \partial_kV &= V G_k +e^G\left(g^{i\bar{j}}\nabla_k G_i G_{\bar{j}}-G_k\right),\\
    \partial_{\bar{l}}\partial_k V &= \partial_{\bar{l}}VG_k+\partial_kV G_{\bar{l}}+V(g_{k\bar{l}}-G_k G_{\bar{l}})+ e^G\left(\nabla_k G_i \nabla_{\bar{l}}G^{i}-R_{k\bar{l}i\bar{j}}G^iG^{\bar{j}}+g_{k\bar{l}}\right), \nonumber \\
   \nabla_l \partial_k V &= \partial_l V G_k + \partial_kV G_l + V\left(\nabla_l G_k - G_lG_k\right)+e^{G}\left(\nabla_l G_k + \nabla_k G_l + G^i \nabla_l \nabla_k G_i\right),
   \nonumber
\end{align}
which, once evaluated at the background with $V= 0$ and $\partial_lV= 0$, yields the matrix elements of the bosonic mass matrix \eqref{bosonicM2}, namely
\begin{equation}
    M^2_{i\bar{j}} = \left\langle \left(\nabla_i G_k \nabla_{\bar{j}}G^k - R_{i\bar{j}k\bar{l}}G^k G^{\bar{l}}+K_{i\bar{j}} \right)e^{G}\right\rangle, \quad M^2_{ij} = \left\langle \left(\nabla_i G_j + \nabla_j G_i+ {G^i\nabla_l \nabla_k G_i}\right)e^{G}\right\rangle,
\end{equation}
where the last term in $M_{ij}^2$ is absent in \cite{Wess:1992cp}. Note that the indices here correspond to field space indices, and  the field space of chiral multiplets is a K\"ahler space. Hence, only the components
\begin{equation}
    \Gamma_{ik}^l = g^{l\bar{j}}\partial_k g_{i\bar{j}},\quad R_{i\bar{j}k\bar{l}} = g_{m\bar{j}}\partial_{\bar{l}}\Gamma_{ik}^{m},\quad R_{k\bar{l}} = g^{i\bar{j}}R_{i\bar{j}k\bar{l}}, 
\end{equation}
together with the ones connected to them by symmetry, are non-zero. The metric $g_{i\bar{j}}$ is the K\"ahler metric, $g_{i\bar{j}}=K_{i\bar{j}}$. The brackets in the matrices above represent evaluation at the background field configuration.

Now that, once we have the mass matrices, we can in principle calculate the supertraces in \eqref{1loopeffectivepot}. In our case, the first term vanishes 
\begin{equation}
    \text{Str}\; \mathbf{1} = \sum_j(-1)^{2j}n_j = 0,
\end{equation}
because supergravity has the same number of propagating bosonic and fermionic degrees of freedom: the scalar 
contributions cancel with the spin-1/2 ones, while the gravitino contribution cancels the graviton contribution. 
To evaluate the second term in \eqref{1loopeffectivepot} we need to calculate
\begin{equation}
    \text{Str} M^2 = \text{tr}M_0^2 -2 \text{tr}M_{1/2}^2 - 4 m_{{\tilde{g}}}^2.
\end{equation}
The result can be found in \cite{Wess:1992cp}:
\begin{equation}\label{strM2}
    \text{Str} M^2 \, = \, 2 ( (n-1)- R_{k\bar{l}}G^k G^{\bar{l}} ) e^G ,
\end{equation}
where $n$ is the number of chiral superfields (two, in our $(S,T)$ model). The expression above is valid for any 
chiral supergravity model, so we simply need to specialize it to our case. Our generalized potential is given by
\begin{subequations}
\begin{align}
    G &= -\ln(S+\bar{S}) - 3\ln(T+\bar{T}) +\ln(W_0 - A e^{-a_0S})+\ln(W_0 - Ae^{-a_0\bar{S}})\\
    G_T &= -\frac{3}{(T+\bar{T})}\\
    G_S & = \frac{1}{(S+\bar{S})}\frac{-W_0+A e^{-a_0S}(1+a_0(S+\bar{S}))}{W_0-Ae^{-a_0 S}}.
\end{align}
\end{subequations}
Using that
\begin{equation}
    R_{k\bar{l}} = \partial_{\bar{l}}(K^{m\bar{n}}\partial_k K_{m\bar{n}}),
\end{equation}
one can show that
\begin{equation}
    R_{S\bar{S}} = \frac{2}{(S+\bar{S})^2}, \quad R_{S\bar{T}} = 0 = R_{T\bar{S}}, \quad R_{T\bar{T}} = \frac{2}{(T+\bar{T})^2},
\end{equation}
such that (note $G^k = K^{k\bar{l}}G_{\bar{l}}$)
\begin{align}
    R_{k\bar{l}}G^k G^{\bar{l}} &= R_{T\bar{T}}G^TG^{\bar{T}}+R_{S\bar{S}}G^SG^{\bar{S}}\nonumber\\
    &= 2+ 2\left|\frac{W_0-Ae^{-a_0S}(1+a_0(S+\bar{S}))}{W_0 - Ae^{-a_0S}}\right|^2.
\end{align}
Plugging this result into equation \eqref{strM2}, we find that
\begin{equation}\label{strM2ourmodel}
    \text{Str} M^2 = 2\left\langle \left(-|W|^2-2|W-Aa_0e^{-a_0S}(S+\bar{S})|^2\right)e^K\right\rangle = -2 \left\langle e^K|W|^2\right\rangle = -2m_{{\tilde{g}}}^2\,,
\end{equation}
where in the second equality we used the fact that, in our model, $\langle D_S W \rangle=0$. 
Note that because of the no-scale structure of the K\"ahler potential we have that
\begin{equation}
    K^{T\bar{T}}D_T W\overline{D_T W} = -3|W|^2 \, .
\end{equation}

To calculate the last term in \eqref{1loopeffectivepot}, we need to know the explicit components of the 4th and 2nd powers of the mass-matrices. To get that, we first compute the non-vanishing connection components
\begin{equation}
    \Gamma_{SS}^S = -\frac{2}{S+\bar{S}} = \Gamma_{\bar{S}\bar{S}}^{\bar{S}} , \quad \Gamma_{TT}^T =-\frac{2}{T+\bar{T}} =\Gamma_{\bar{T}\bar{T}}^{\bar{T}},
\end{equation}
such that
\begin{equation}
    \nabla_S G_S = -G_S(G_S-a_0)+\frac{a_0}{S+\bar{S}}, \quad \nabla_T G_T = -\frac{3}{(T+\bar{T})^2} = -\frac{1}{3}G_TG_{\bar{T}}, \quad \nabla_S G_T = 0 = \nabla_T G_S.
\end{equation}
The non-zero components of the Riemann tensor are
\begin{equation}
    R_{S\bar{S}S\bar{S}} = \frac{2}{(S+\bar{S})^4}, \quad R_{T\bar{T}T\bar{T}} = \frac{6}{(T+\bar{T})^4},
\end{equation}
and their complex conjugates. Note that the simple structure of these components comes from the simple form of the K\"ahler metric, which is diagonal and has $S\bar{S}$-component independent of $T$ and $T\bar{T}$-component independent of $S$.

Using the results above, one can show that the non-vanishing components of the scalar mass matrix are\footnote{If we do not include the last term in $M^2_{ij}$, the component $M_{TT}^2$ does not vanish.}
\begin{equation}
    (M_0^2)_{SS} = \frac{a_0}{S_0}m_{{\tilde{g}}}^2, \quad 
    (M_0^2)_{S\bar{S}} = \left(a_0^2 +\frac{1}{4S_0^2}\right)m_{{\tilde{g}}}^2,
\end{equation}
and their complex conjugates. The non-vanishing component of the spin-1/2 mass-matrix is
\begin{equation}
    (M_{1/2}^2)_{S\bar{S}} = a_0^2 m_{{\tilde{g}}}^2.
\end{equation}
In matrix form,
\begin{equation}
    M_0^2 = m_{{\tilde{g}}}^2\begin{pmatrix}
    a_0^2 +\frac{1}{4S_0^2} & 0 & \frac{a_0}{S_0} & 0\\
    0 & 0 & 0 & 0\\
    \frac{a_0}{S_0} & 0 & a_0^2 +\frac{1}{4S_0^2} & 0\\
    0 & 0 & 0 & 0\\ 
    \end{pmatrix}, \quad M_{1/2}^2 = m_{{\tilde{g}}}^2\begin{pmatrix}
    a_0^2 &  0\\
    0 & 0\\
    \end{pmatrix}.
\end{equation}
Because the K\"ahler field space has non-vanishing curvature, in order to make sense of the $\ln$ of the mass-matrix, 
we write it as the $\ln$ of an operator. This is accomplished by writing (for the fermionic trace)
\begin{align}
    \text{tr}M^4_{1/2}\ln \left(\frac{M^2_{1/2}}{\bar{\Lambda}^2}\right) &= (M_{1/2}^4)_{i\bar{k}}g^{l\bar{k}}\left[\ln \left(\frac{M^2_{1/2}}{\bar{\Lambda}^2}\right)\right]_{l\bar{j}}g^{i\bar{j}}\nonumber\\
    & = (M^2_{1/2}\cdot g^{-1}\cdot M^2_{1/2}\cdot g^{-1})_i^{\;l}\left[\ln \left(\frac{M^2_{1/2}\cdot g^{-1}}{\bar{\Lambda}^2}\right)\right]_{l}^{\;i}.
\end{align}
Hence the relevant operator for computing the fermionic part of the supertrace is
\begin{equation}
    M^2_{1/2}\cdot g^{-1} = m_{{\tilde{g}}}^2\begin{pmatrix}
    (2S_0 a_0)^2 & 0\\
    0 & 0\\
    \end{pmatrix},
\end{equation}
where we have evaluated the inverse K\"ahler metric at the background configuration. Since the operator is diagonal, we get
\begin{equation}
    \text{tr}M^4_{1/2}\ln \left(\frac{M^2_{1/2}}{\bar{\Lambda}^2}\right) = (2S_0 a_0)^4 m_{\tilde{g}}^4 \ln \left(\frac{(2S_0 a_0)^2m_{{\tilde{g}}}^2}{\bar{\Lambda}^2}\right).
\end{equation}
Note that the zero eigenvalue of $M_{1/2}^2\cdot g^{-1}$ does not harm the final result because $x^2\ln x \to 0$ as $x \to 0$.

For the bosonic mass-matrix, the relevant operator will turn out to be
\begin{equation}
    M_0^2\otimes g^{-1} = \begin{pmatrix}
    M^2_{i\bar{j}}g^{\bar{j}k} & M_{ij}^2g^{j\bar{k}} \\
    M^2_{\bar{i}\bar{j}}g^{\bar{j}k} & M^2_{\bar{i}j}g^{j\bar{k}}
    \end{pmatrix} = \begin{pmatrix}
    (M^2)_i^{\;k} & (M^2)_i^{\;\bar{k}}\\
    (M^2)_{\bar{i}}^{\;k} & (M^2)_{\bar{i}}^{\;\bar{k}} \end{pmatrix} = m_{\tilde{g}}^2\begin{pmatrix}
    1+(2S_0a_0)^2 & 0 & 4S_0 a_0 & 0 \\
    0 & 0 & 0 & 0 \\
    4S_0 a_0 & 0 & 1+(2S_0 a_0)^2 & 0\\
    0 & 0 & 0 & 0
    \end{pmatrix},
\end{equation}
which is not diagonal. Its eigenvectors are
\begin{equation}
    v_1 = \begin{pmatrix}
    0\\
    0\\
    0\\
    1
    \end{pmatrix},\quad
    v_2 = \begin{pmatrix}
    0\\
    1\\
    0\\
    0
    \end{pmatrix}, \quad
    v_3 = \begin{pmatrix}
    -1\\
    0\\
    1\\
    0
    \end{pmatrix}, \quad
    v_4 = \begin{pmatrix}
    1\\
    0\\
    1\\
    0
    \end{pmatrix},
\end{equation}
and its eigenvalues are given by
\begin{equation}
    \lambda_1 = 0 = \lambda_2, \quad \lambda_3 = (1-2a_0S_0)^2m_{\tilde{g}}^2, \quad \lambda_4 = (1+2a_0S_0)^2m_{\tilde{g}}^2. 
\end{equation}
The bosonic trace is then found to be
\begin{align}
    \text{tr}M_0^4\ln\left(\frac{M_0^2}{\bar{\Lambda}^2}\right) &= \sum_i \lambda_i^2 \ln\left(\frac{\lambda_i}{\bar{\Lambda}^2}\right)\\
    &= (1-2a_0S_0)^4 m^4_{\tilde{g}} \ln\left(\frac{(1-2a_0S_0)^2m_{\tilde{g}}^2}{\bar{\Lambda}^2}\right)+(1+2a_0S_0)^4 m^4_{\tilde{g}} \ln\left(\frac{(1+2a_0S_0)^2m_{\tilde{g}}^2}{\bar{\Lambda}^2}\right),
\end{align}
and the supertrace in the last term of \eqref{1loopeffectivepot} is
\begin{align}\label{strM4ourmodel}
    \text{Str}M^4\ln\left(\frac{M^2}{\bar{\Lambda}^2}\right) &= (1-2a_0S_0)^4 m^4_{\tilde{g}} \ln\left(\frac{(1-2a_0S_0)^2m_{\tilde{g}}^2}{\bar{\Lambda}^2}\right)+(1+2a_0S_0)^4 m^4_{\tilde{g}} \ln\left(\frac{(1+2a_0S_0)^2m_{\tilde{g}}^2}{\bar{\Lambda}^2}\right)-\nonumber\\
    &-2(2a_0S_0)^4m_{\tilde{g}}^4 \ln\left(\frac{(2a_0S_0)^2m_{\tilde{g}}^2}{\bar{\Lambda}^2}\right).
\end{align}
If $a_0S_0$ is large (as we require for phenomenological reasons - see Section III) then we have some cancellations between the positive and negative terms above:
\begin{equation}
    \text{Str}M^4 \ln\left(\frac{M^2}{\bar{\Lambda}^2}\right) = \left(\frac{25}{3}+14(2a_0S_0)^2\right)m^4_{\tilde{g}} +\left(2+12(2a_0S_0)^2\right)m^4_{\tilde{g}} \ln\left((2a_0S_0)^2\frac{m^2_{\tilde{g}}}{\bar{\Lambda}^2}\right)+\mathcal{O}\left(m^4_{\tilde{g}}(2a_0S_0)^{-2}\right)
\end{equation}

Summarizing, using the supertraces \eqref{strM2ourmodel} and \eqref{strM4ourmodel} and assuming the weakly coupled limit, we find
\begin{equation}
    V_{1l} \approx \frac{\Lambda_c^2}{16\pi^2} m_{\tilde{g}}^2 - \frac{1}{64\pi^2}\left(\frac{25}{3}+14(2a_0S_0)^2\right)m^4_{\tilde{g}}- \frac{1}{64\pi^2}\left(2+12(2a_0S_0)^2\right)m^4_{\tilde{g}} \ln\left((2a_0S_0)^2\frac{m^2_{\tilde{g}}}{\bar{\Lambda}^2}\right).
\end{equation}

Using that $T+\bar{T} = 2 e^{\sqrt{2/3}\psi}$ in the expression for the gravitino mass shows that  (restoring the factors of Planck mass)
\begin{equation}
    m_{\tilde{g}}^2 = \left\langle e^{ \kappa^2 K}|W|^2 \right\rangle = \frac{\left(2Aa_0S_0e^{-a_0S_0}\right)^2}{2S_0(T+\bar{T})^3} = \frac{A^2a_0(2a_0S_0)e^{-2a_0S_0}}{8}e^{-\sqrt{6}\psi}=: \gamma^2 e^{-\sqrt{6}\psi},
\end{equation}
and plugging this result into the one-loop correction yields
\begin{equation}
    V_{1l} \approx \frac{\Lambda_c^2\gamma^2}{16\pi^2}e^{-\sqrt{6}\psi} - \frac{\gamma^4}{64\pi^2}\left[\frac{25}{3}+14(2a_0S_0)^2+\left(2+12(2a_0S_0)^2\right)\left(-\sqrt{6}\psi+\ln\left(\frac{(2a_0S_0)^2\gamma^2}{\bar{\Lambda}^2}\right)\right)\right]e^{-2\sqrt{6}\psi}.
\end{equation}
For large values of $\psi$, the first term dominates.

The leading term in the one-loop potential can also be calculated for the general case of multiple running K\"ahler moduli. Assuming the stabilized moduli to satisfy $\langle G_i\rangle$ = 0 with $\langle V \rangle =0$, the contribution of $n-1$ running moduli and the axion-dilaton to the supertrace of the squared mass-matrix is 
\begin{equation}
    \langle \text{Str}M^2 \rangle = 2(3-2n)m^2_{\tilde{g}},
\end{equation}
where the previous results are recovered when $n=2$, corresponding to $S$ and one running K\"ahler modulus, but with $T$ not necessarily equal to the overall volume. In this case, we get the same dominant contribution for the one-loop potential,
\begin{equation}
    V_{1l} \approx \frac{\Lambda_c^2}{16\pi^2}m^2_{\tilde{g}} = \frac{\Lambda_c^2 \gamma^2}{16\pi^2} \frac{1}{\langle\mathcal{V}\rangle},
\end{equation}
with a possible different relation between $\mathcal{V}$ and $T$ compared to the case when $T$ is the overall volume.

In the expressions above, $\Lambda_c$ is a cutoff scale corresponding to the energy scale at which the four-dimensional effective description breaks down. From a bottom-up perspective, it is clear that this happens once we start exciting Kaluza-Klein modes, for after that one cannot trust the $\mathcal{N}$=1 $d=4$ description. So we identify,
\begin{equation}
    \Lambda_c = m_{\text{KK}} = \frac{M_s}{R},
\end{equation}
where $R\sim \mathcal{V}^{1/6}$ is the characteristic radius of the internal manifold (in string units). 


This correction to the final potential should have a ten-dimensional counterpart. In particular, loop corrections in string theory correspond to $g_s$ corrections to the effective ten-dimensional action. In fact, the K\"ahler potential is not protected by nonrenormalization theorems and receives perturbative corrections (see e.g. \cite{Cicoli:2007xp} for a check that corrections to the K\"ahler potential match the four-dimensional effective potential calculated from the Coleman-Weinberg formula). Moreover, in \cite{Burgess:2020qsc}, a formal argument based on supersymmetry and scale symmetry of the ten-dimensional supergravity model predicts the existence of corrections to $K$. 

\section*{Appendix C: Contribution to the Potential from Corrections to the K\"ahler Potential}

In this appendix (working again in Planck units) we consider perturbative corrections to the K\"ahler potential and demonstrate that they also give contributions to the potential of runaway form. For concreteness, let us consider a supergravity model with several K\"ahler moduli,
\begin{equation}
    K = - \ln(S+\bar{S}) - \ln \mathcal{V}, \quad \mathcal{V} = \frac{1}{6}d_{ijk}t^it^jt^k = c_{ijk}(T_i+\bar{T}_j)(T_j+\bar{T}_j)(T_k+\bar{T}_k),
\end{equation}
where $\mathcal{V}$ is the Einstein frame volume of the extra dimensions in string units, $d_{ijk} =6\times 2^3 c_{ijk}$ are intersection numbers of the internal manifold and $T_i$ are the K\"ahler moduli. Perturbative corrections to the K\"ahler potential will have the form
\be
\delta K = c\mathcal{V}^{-n} \, , 
\ee
where the value of $n$ depends on which kind of correction we consider. From \cite{Cicoli:2013rwa}, we have $n = 2/3$ for $\mathcal{O}(\alpha'^2)$ corrections, and $n=1$ for $\mathcal{O}(\alpha'^3)$ corrections, where the latter is absent in the case of the standard embedding.  

Let us consider the leading correction to the four-dimensional potential induced by $\delta K$. We have
\begin{equation}
    \delta V = \delta K V + e^{K}\left[\delta K^{i\bar{j}}D_iW D_{\bar{j}}\overline{W} + K^{i\bar{j}}\delta(D_j W)D_{\bar{j}}\overline{W}+ K^{i\bar{j}}D_iW \delta(D_{\bar{j}}\overline{W})\right].
\end{equation}
Assuming $W$ to be independent of $T^{i}$, we have $\delta(D_i W) = \delta K_i W$ and
\begin{equation}
    \delta V = \delta K V + e^{K}|W|^2\left[\delta K^{i\bar{j}}K_iK_j+K^{i\bar{j}}(\delta K_i K_{\bar{j}}+ K_i \delta K_{\bar{j}})\right].
\end{equation}
For $K = -\ln(S+\bar{S}) -\ln \mathcal{V} + c \mathcal{V}^{-n}$, we have
\begin{align}
    K_i &= \frac{\partial K}{\partial T^i} = -\frac{3}{2}\frac{\mathcal{V}_i}{\mathcal{V}} - n c \mathcal{V}^{-n}\frac{\mathcal{V}_i}{\mathcal{V}}\,,\\
    K_{i\bar{j}} &= -\frac{3}{2}\left(\frac{\mathcal{V}_{i\bar{j}}}{\mathcal{V}}-\frac{3}{2}\frac{\mathcal{V}_i \mathcal{V}_{\bar{j}}}{\mathcal{V}^2}\right) -n c \mathcal{V}^{-n}\left(\frac{\mathcal{V}_{i\bar{j}}}{\mathcal{V}}-(n+1)\frac{\mathcal{V}_i \mathcal{V}_{\bar{j}}}{\mathcal{V}}\right)\,,
\end{align}
and, to linear order in $c$,
\begin{equation}
    K^{i\bar{j}} = -\frac{2}{3}\mathcal{V}\mathcal{V}^{i\bar{j}} +2 t^i t^j + \frac{4}{9}c n \mathcal{V}^{-n}\left(\mathcal{V}\mathcal{V}^{i\bar{j}}-4\left(n+\frac{1}{4}\right)t^{i}t^{j}\right)\,,
\end{equation}
where $\mathcal{V}^{i\bar{j}}$ is the inverse of $\mathcal{V}_{i\bar{j}} = (1/6)c_{ijk}t^{k}$  and $\mathcal{V}_i = (1/6)c_{ijk}t^jt^k$.

Using these relations, we get
\begin{equation}
    \delta V = c \mathcal{V}^{-n}V - n(n-1)4c\mathcal{V}^{-n}e^{K}|W|^2.
\end{equation}

In our case, we are interested in the corrections around the background with stabilized axio-dilaton, in which $\langle V\rangle = 0$. So,
\begin{equation}
    \langle \delta V \rangle  = -n(n-1)4c \mathcal{V}^{-n}\langle e^{K}|W|^2\rangle = -n(n-1)4c \mathcal{V}^{-n}m_{\psi}^2 \sim \frac{\langle|W|^2\rangle}{2S_0\mathcal{V}^{n+1}}.
\end{equation}
Note that for $n=1$, the leading correction vanishes, which is due to the ``extended no-scale structure" of $K$, as explained in \cite{Cicoli:2007xp}. 

For $n=2/3$ ($\mathcal{O}(\alpha'^2)$ corrections),
\begin{equation}
    \langle \delta V\rangle \sim \frac{\langle |W|^2 \rangle}{2S_0}\frac{1}{\mathcal{V}^{5/3}} = \frac{\langle |W|^2 \rangle}{2S_0}\frac{2^5}{(T+\bar{T})^{5}},
\end{equation}
where in the last line we assumed a single K\"ahler modulus, such that $\mathcal{V}^{1/3} = (T+\bar{T})/2$.

For $n=4/3$, we have
\begin{equation}
    \langle \delta V\rangle \sim \frac{\langle |W|^2 \rangle}{2S_0}\frac{1}{\mathcal{V}^{7/3}} = \frac{\langle |W|^2 \rangle}{2S_0}\frac{2^7}{(T+\bar{T})^{7}} \, .
\end{equation}
In either case, we conclude that corrections to the K\"ahler potential of the form $\delta K = c\mathcal{V}^{-n}$ will produce a term in the potential that has the form
\begin{equation}
    \langle \delta V \rangle \sim \frac{m_\psi^2}{\mathcal{V}^{n}}=\frac{\langle |W|^2\rangle}{2S_0}\frac{1}{\mathcal{V}^{n+1}} = \frac{\langle |W|^2\rangle}{2S_0}\frac{2^{\frac{n+1}{3}}}{(T+\bar{T})^{\frac{n+1}{3}}}\,.
\end{equation}

Note that the overall amplitudes of these $\alpha'$ correction terms are of the same order of magnitude as that of the one loop correction to the potential, but the terms fall off with a higher power of $e^{-\psi}$. Hence, for the large values of the modulus field $\psi$ which we consider, the $\alpha'$ terms are subdominant. Note, in addition, that the $\mathcal{O}(\alpha'^2)$ term vanishes for the standard embedding of the gauge group, and that hence the $\mathcal{O}(\alpha'^3)$ corrections are the leading ones in that case \cite{Cicoli:2013rwa}. This would result in a further suppression of the $\alpha'$ terms compared to the one loop corrections.  

\section*{Appendix D: Determination of the cutoff scale $\Lambda_c$}

The cutoff scale $\Lambda_c$ is the energy scale at which the four-dimensional effective field theory description breaks down. This happens when it becomes possible to start exciting Kaluza-Klein modes. To calculate this scale, consider the relation between the ten-dimensional and the four-dimensional metrics (see Appendix A)
\be
ds^2_{10, S} \, = \, e^{\Phi} g_{\mu \nu} dx^{\mu} dx^{\nu} + e^{\Psi} h_{mn} dy^m dy^n \, .
\ee
From the relation between these metrics, the Kaluza-Klein length scale $R_{KK}$ is given by
\be
e^{\Phi} R_{KK}^2 \, \sim \, e^{\Psi} V_6^{1/3} \, ,
\ee
where ${\tilde{V}}_6$ gives the volume of the fiducial metric $h_{mn}$ in string units, i.e.
\be
V_6 \, = \, {\tilde{V}}_6 m_s^{- 6} \, .
\ee
The dimensionless number ${\tilde{V}}_6$ gives the relationship between the four dimensional Planck mass and the string scale (which, up to a numerical constant which depends on the specific string theory we are considering, is equal to the ten-dimensional Planck mass):
\be
m_{pl} \, = \, m_s ({\tilde{V}}_6)^{1/2} \, .
\ee
Combining these equations then yields
\be
\Lambda_c^2 \, \sim \left( \frac{1}{R_{KK}} \right)^2 \, \sim \, e^{\Phi} 
\left( \frac{m_s}{m_{pl}} \right)^{8/3} e^{- \Psi}  m_{pl}^2 \, .
\ee

\end{document}